THIS RESPONSE CONTAINS TWO PARTS:

The first part is Response to Perk's Comment, the second part is Reply to Perk's Rejoinder, focusing on singularities at/near infinite temperature, also with additional replies 1 & 2

# PART I

## Response to 'Comment on 'Conjectures on exact solution of three-dimensional (3D) simple orthorhombic Ising lattices'' by Perk

Z.D. Zhang*

*Shenyang National Laboratory for Materials Science, Institute of Metal Research and International Centre for Materials Physics, Chinese Academy of Sciences, 72 Wenhua Road, Shenyang, 110016, P.R. China*

The error of eq. (15b) in my article [Z.D. Zhang, Phil. Mag. **87**, 5309 (2007) and also see arXiv: 0705.1045] in the application of the Jordan-Wigner transformation does not affect the validity of the putative exact solution, since the solution is not derived directly from it. Other objections of Perk's Comment [J.H.H. Perk, Phil. Mag. **89**, (2009) 761, also see arXiv:0811.1802v2] are the same as those in Wu et al.'s Comments [F.Y. Wu et al., Phil. Mag. **88**, (2008) 3093; 3103], which do not stand on solid ground and have been rejected in my previous Response [Z.D. Zhang, Phil. Mag. **88**, (2008) 3097]. The conjectured solution can be utilized to understand critical phenomena in various systems, while the conjectures are open to rigorous prove.

*Email: zdzhang@imr.ac.cn

The present paper is a Response to Perk's Comment [1] on the conjectured solution of the three-dimensional (3D) Ising model [2]. Firstly, I should like to thank Professor Perk for pointing out the error of eq. (15b) in [2] in the application of the Jordan-Wigner transformation, which should be corrected as eq. (3) of [1].[1] However, although this error is the same as Maddox's [4], the essential difference is that my putative solution in [2] is obtained by introducing two conjectures dealing with the topologic problem in the 3D Ising model, and is not derived directly from the erroneous equation. Thus the error does not affect the validity of the putative exact solution. When I wrote [2], I thought that the topologic troubles were due to the U factors in eq. (15). According to Lou and Wu [3] and following my discussions with Perk, the appearance of the high-order terms in eq. (3) of [1] and the corresponding exponential factors of the transfer matrix (not the U factors in eq. (15) of [2]) is the root of the difficulties with the 3D Ising model. The *U* factors come from the periodic boundary conditions, but disappear for open boundary condition. It is clear now that there is no need to remove the U factors at the boundary by some topologic trick, but the high-order 'internal' factors might need closer attention. Since there is an 'internal' factor for each j (j runs from 1 to *nl* in [1], corresponding to (r, s) running from (1, 1) to (*n, l*) in [2]), the number of the 'internal' factors is in the order of *nl* more than that of the *U* factors. These 'internal' factors raise more difficulties since they do not commute with the rest in the transfer matrix (e.g., the product of the factors like $e^{\frac{1}{2}\alpha\Gamma}$ [3]) and one does not have representations of the rotation group. So that one cannot continue as Kaufman did [5]. Although the situation becomes more complicated, the same conjectures can still be made, with the motivation for the conjectures being only slightly different with what I had in [2]. Suppose that we are given a 3D manifold bounding a 4D manifold [6,7]. It might be possible to attach an "internal" space on every point of the 3D lattice to provide with some operators to allow these "internal" factors to commute with the transfer matrix. In this sense, we add an extra dimension with an additional rotation as a kind of boundary condition as what I conjectured in my original paper [2]. Then we might solve simultaneously the topologic problem in eq. (15) as a whole regarding to its non-local behavior, no matter how complicated it is.

---

[1] Lou and Wu [3] also apply correctly an equivalent equation. An error in eq. (16) of [2] should be corrected as eq. (1) of [1]. A factor of 2 should be added in all the exponentials on the right hand of eq. (28) and also the **Conjecture 2** on page 5321 of [2] (e.g. $e^{i\frac{2\pi}{n}}, \ldots e^{i\frac{2t_x\pi}{n}}$, etc.).

Istrail showed that the essential ingredient in the NP-completeness of the Ising model is nonplanarity [8], which indicates also that the origin of difficulties is topologic. As discussed on page 5393 of [2], the NP-completeness only prevents algorithms from solving all instances of the problem in polynomial time [9,10]. Such NP-completeness from the point view of computer sciences cannot be fully used to judge the advances in mathematics that are needed to uncover the exact solution. Furthermore, as Istrail and Cipra claimed [8-10], there exists the possibility for exact answers in the ferromagnetic 3D Ising model dealt with in [2]. As discussed above, the main difficulties caused by these high-order terms are topologic [1,8-10]. Therefore, the conjectures of introducing the fourth dimension [2], which serve for dealing with the topologic problem in the 3D Ising model, are still meaningful, and open to be proved rigorously (with, however, the new form of eq. (15b) for the matrix V [2], thanks to Professor Perk).

Other objections in [1], concentrating on the low- and high-temperature expansions and the different choices of the weight functions, are all the same as those in recent Comments by Wu et al. [11,12], which have been rejected in my previous Response [13]. The only exception is that the literatures referred to in [1] for rigorously proving the convergence of the high-temperature series [14-18] are different with those [19-22] in [11]. As remarked in [1], the proof of [14-18] is based on the proof of Gallavotti and Miracle-Solé [14]. However, just below eq. (5) of [14], the authors put for convenience $\beta = 1/(k_B T) = 1$, which is inconsistent with $\beta = 0$ for infinite temperature. Some important conditions for Theorems in [14] are not valid for $\beta = 0$. For instance, the condition for ii), iii) and iv) of Theorem 1, Theorems 2 and 3, will be invalid if $\beta = 0$ is put into eq. (24) of [14]. One may argue that infinite temperature just requires that all interaction energies equal to zero. But in this case, this condition is invalid still and, moreover, one should face a change of all the interaction energies from zero to non-zero at/near $\beta = 0$. Such change results in an intrinsic change of the geometrical (topologic) structure in the 3D Ising interaction system as revealed in [13]. As has been already pointed out in [13], Lebowitz and Penrose [15] and Griffiths [17] distinguished $\beta > 0$ and $\beta = 0$, and started with the condition $\beta > 0$ to prove their theorems. The basic difficulty of these well-known theorems originates from a fact that a phase transition may occur at $\beta = 0$ according to the Yang–Lee theorems [23,24].

Everyone has been brought to a situation in which it is impossible to satisfy the opposite wishes: being convergent as an exact solution is, while it must agree exactly with a divergent series. As remarked in [1], the low-temperature series of my putative exact solution has a finite radius of convergence up to its critical point. So it is to be expected that it does not reproduce term by term the well-known low-temperature series that is divergent. The lack of information of the global behaviours of the 3D Ising system is the root of such divergence in the well-known low-temperature series. The troubles with it may originate from some difficulties in the foundation of statistical mechanics [25-29].

In [13], I indicated the necessity of introducing a (3+1)–dimensional framework for dealing with the 3D Ising model and discussed briefly the physics beyond the extra dimension. According to [12],[2] it may be profitable to inspect further the mathematical basis of statistical mechanics, i.e., the *ergodic hypothesis* and the *mixing hypothesis* [25-31]. The *ergodic hypothesis* has been proved to be one of the most difficult problems and its proof under fairly general conditions is lacking [25-31]. For the *mixing hypothesis* which is stronger than *the ergodic hypothesis*, talking about a distribution of points on the surface $\Gamma$ (E), one is no longer discussing a single system, and mixing is irrelevant for a truly isolated system [25]. In statistical mechanics, one simply assumes that the time average can be replaced by the ensemble average [25-31]. Actually, most systems studied in statistical mechanics are not ergodic [29-31]. It is my understanding that the lack of ergodicity of the 3D Ising model would lead to that the time average being different from the ensemble average, which may not contain complete information of the system. Neglecting the difference between the two averages may work well in other models with dimensions D ≠ 3, but cause serious troubles in the 3D Ising system because of its global topologic behaviour and geometrical structure [2,13]. Since the well-known low- and high-temperature series of the 3D Ising model might not account properly for the time average of the system, they might be invalid at finite temperatures. In my view, it is unjustified to reply upon successes of statistical mechanics to dismiss questions regarding its foundation.

---

[2] I take this opportunity to reply briefly the last sentence in Wu et al.'s Rejoinder [12] to my Response [13].

In summary, the error in [2] should be corrected as Perk suggested in [1], but it does not affect the validity of the putative exact solution, which is not derived directly from the erroneous equation. All these well-known theorems in [14-22] are proved only for $\beta > 0$, not for infinite temperature. Other objections in [1] and also those in [11,12] do not stand on solid ground. The conjectured solution can be utilized to understand critical phenomena in various systems [32,33], while the conjectures are still open to rigorous prove.

**Acknowledgements**

The author appreciates the supports of the National Natural Science Foundation of China (under grant numbers 50831006 and 10674139). I am grateful to Prof. Dr. Jacques H.H. Perk for helpful discussion via e-mail exchanges.

# PART II

# Singularities at/near infinite temperature: Reply to Perk's Rejoinder on 'Conjectures on exact solution of three-dimensional (3D) simple orthorhombic Ising lattices'

Z.D. Zhang

*Shenyang National Laboratory for Materials Science, Institute of Metal Research and International Centre for Materials Physics, Chinese Academy of Sciences, 72 Wenhua Road, Shenyang, 110016, P.R. China*

This is a Reply to the Rejoinder (J.H.H. Perk, Phil. Mag. **89**, (2009) 761, arXiv:0901.2935) to the Response (Z.D. Zhang, Phil. Mag. **89**, (2009) 765, arXiv:0812.0194) on Perk's Comment (J.H.H. Perk, Phil. Mag. **89**, (2009) 769, arXiv:0811.1802) on 'Conjectures on exact solution of three-dimensional (3D) simple orthorhombic' (Z.D. Zhang, Phil. Mag. **87**, (2007) p.5309, arXiv:0705.1045). It is shown that the basis of the objections in Perk's Rejoinder (arXiv:0901.2935), with respect to singularities at/near infinite temperature, is based on an error that mixes the concepts $T \to \infty$ and $T = \infty$ (i.e., $\beta \to 0$ and $\beta = 0$, with $\beta \equiv (k_B T)^{-1}$). It is shown that the reduced free energy per site $\beta f$ can be used only for finite temperatures ($\beta > 0$), not for "exactly" infinite temperature ($\beta = 0$). Thus, the convergence of the well-known high-temperature series has not been rigorously proved for $\beta = 0$. Furthermore, at the thermodynamic limit $N \to \infty$, besides infinite-temperature zeros of Z at $z = -1$ for $H = \pm i\infty$ in the limit $\beta \to 0$, there exists another singularity at $z = 1$ for the partition function as well as the high-temperature series, which is usually concealed in literature by setting $Z^{1/N}$ and dividing the total free energy F by N (equally, disregarding the singularity of zeros of $Z^{-1}$). Therefore, the well-known high-temperature series cannot serve as a standard for judging the putative exact solution of the 3D Ising model. The objections in Perk's Rejoinder (arXiv:0901.2935) are thoroughly disproved.

After publication of the conjectured exact solution of three-dimensional (3D) simple orthorhombic Ising lattices [1], there have been two rounds of exchanges of Comments/Responses/Rejoinders [2-7]. After all discussions in [2-7], it seems that the only key issue left is that of singularities at/near infinite temperature. Both groups of authors (Wu et al. and Perk) of the Comments/Rejoinders [2,4,5,7] insist that the procedure in [1] is wrong, because the conjectured free energy can fit well with the well-known high-temperature series only at/near infinite temperature, as the convergence of the high-temperature series has been rigorously proved in [8-16]. The proposals of this Reply are to discuss in detail the singularities at/near infinite temperature and also to point out that there is an error in Perk's Rejoinder [7], which is the basis of the arguments with respect to the high-temperature series in [2,4,5,7].

As noted in the previous Responses [3,6], all the rigorous theorems in [8-16] have been proved only for $\beta \equiv 1/(k_BT) > 0$, i.e., $T < \infty$. Exactly infinite temperature has been never touched in these theorems, since there is a possibility of the existence of a phase transition at $\beta = 0$, according to the condition of $z \equiv \exp(-2\beta H) = 1$ in the Yang-Lee Theorem [17,18]. There are three possibilities for the existence of a phase transition: 1) $H = 0$, $\beta \neq 0$; 2) $H \neq 0$, $\beta = 0$; 3) $H = 0$, $\beta = 0$. This point of $\beta = 0$ has been avoided during the procedure of rigorously proving these theorems in [8-16]. The difficulty for a rigorous prove including $\beta = 0$ is due to the fact that there is no general reason to expect a series expansion of *p* or *n* in powers of *β* to converge (p. 102 of [9]), since *β* = 0 lies at the boundary of the region E of ($\beta$, z) space. The difficulty has been bypassed by using the dimensionless parameters $K_i = \beta J_i$, (i = 1,2,3) and h = βH and setting β = 1 during the procedure.

Let us start from the initial point of the problem to discuss in detail the origin of the singularities at/near infinite temperature. The total free energy of the system is: $F = U - TS = -k_B T \ln Z$. The singularities in the free energy and other thermodynamic consequences (such as the entropy, the internal energy, the specific heat, the spontaneous magnetization, etc) originate from the singularities of the partition function Z. This is why Yang and Lee discuss the phase transition by evaluating the distribution of roots of the grand partition function (i.e., Z = 0) in their general theory [17,18]. In order to describe infinite systems, one usually normalizes the extensive variables that are homogeneous of degree one in the volume, by the volume **V** (or the number of particles N), keeps the density (i. e. the number of particles per volume) fixed and takes the limit for V (or N) tending to infinity. In this sense, one usually defines the thermodynamic limit (N → ∞) for the free energy per site f by $f = F/N = -k_B T \ln \lambda$ with $\lambda = Z^{1/N}$. By such a procedure, it is expected that one can establish the fact that f converges uniformly to its common limit as N → ∞, namely, it is performed with an assumption (or expectation) that f is finite [17-20]. In this way, one can easily avoid to deal with the total free energy $F = -Nk_B T \ln \lambda$ of the system, which shows singularities at any temperature as N → ∞ and if $\ln \lambda$ is finite. However, it is clearly seen that at infinite temperature (T = ∞), there still exists a singularity in the free energy per site f that is equal to negative infinite in the case that $\ln \lambda$ is positive and finite. Actually, $f = -k_B T \ln \lambda = -k_B \ln \lambda^T$, in both forms, f is equal to negative infinite at T = ∞. Using the value of the 3D Ising model λ = 2, one easily finds that $f = -k_B T \ln 2 = -k_B \ln 2^T$ has a singularity at T = ∞. This is inconsistent with the assumption for the definition of the free energy per site f, and therefore, it loses physical significance at T = ∞. It is clear that one has to face directly the total free energy F to study the singularities of the system at T = ∞.

One may argue that such singularities of the whole system are not of physical significance, which should be removed by using the reduced free energy per site βf. As stated in Perk's Rejoinder [7], the reduced free energy per site βf is often rewritten as βf = ϕ({$K_i$}, h) = ϕ({β$J_i$}, βH) with some function ϕ. But the error in [7] is easily seen as follows: Setting β = 1 equalizes to T = 1/$k_B$ ≠ ∞. Therefore, the necessary and sufficient condition for using the dimensionless parameters $K_i$ = β$J_i$, (i = 1,2,3) and h = βH and setting β = 1 is β ≠ 0. Thus, setting β = 1 is loss of generality for β = 0, and the replacements $J_i$ → β$J_i$, H → βH and f → βf are validated only for β → 0. All discussions in the Perk's Rejoinder [7] are only based on the limit β → 0, but not 'exactly' on infinite temperature (T = ∞, β = 0). Thus, the convergence of the well-known high-temperature series for the 3D Ising model has not been rigorously proved for β = 0.

The total free energy of the system can also be written as $F = k_B T \ln Z^{-1}$. Therefore, besides the roots of the partition function Z, one should also discuss the roots of $Z^{-1}$. Writing z ≡ exp(−2βH) and keeping βH fixed in the limit β → 0, the partition function of an arbitrary lattice with N sites for the Ising model becomes Z = $(z^{1/2}+z^{-1/2})^N$ [7]. It is easily seen that $z^{1/2} + z^{-1/2} > 1$ satisfies the condition for the zeros of the reciprocal of the partition function, i.e., $Z^{-1} = (z^{1/2} + z^{-1/2})^{-N}$. So, the infinite-temperature zeros of $Z^{-1}$, i.e., $Z^{-1} \to 0$, occur at $z = 1$ as $N \to \infty$, $Z = 2^N \to \infty$. Or more explicitly speaking, the zeros are located at $\beta = 0$, $z = 1$. The discussion above can be supported by the fact that the singularity behavior of the logarithmic function ln x in the two cases of $x = 0$ and $x = \infty$ correspond to those in logarithmic function ln y with $y = 1/x$ in two cases of $y = \infty$ and $y = 0$, respectively. It indicates clearly that both singularities at the two limits of $Z = 0$ and $Z = \infty$ are actually the same, except for a minus sign, and considerable interest should be paid to both of them.

From the Yang-Lee Theorem [17,18] and the findings above, in the 3D Ising model there indeed exist three singularities: 1) $H = 0$, $\beta = \beta_c$; 2) $H = \pm i\infty$, $\beta \to 0$; 3) $H = 0$, $\beta = 0$. The 3D Ising system experiences a change from a 'non-interaction' state at $\beta = 0$ to an interacting state at $\beta > 0$. This change of the states just likes that there is a 'switch' turning off/on all the interactions at/near infinite temperature, resulting in the change of the topologic structures and the corresponding phase factors [1,3,6].

In summary, the procedure of using the dimensionless parameters $K_i = \beta J_i$, ($i = 1,2,3$) and $h = \beta H$ and setting $\beta = 1$ for rigorously proving the analytic behavior of the free energy can be employed only for $\beta > 0$, not $\beta = 0$. There is an error in Perk's Rejoinder [7], which mixes the concepts of $T \to \infty$ and $T = \infty$ (i.e., $\beta \to 0$ and $\beta = 0$). Besides the singularity at $z = -1$, there is another singularity at $z = 1$ of the partition function as well as the high-temperature series at the thermodynamic limit $N \to \infty$. The latter singularity may not cause problems in dimensions $D \neq 3$, but does cause serious troubles in 3D. This is usually concealed in literatures by setting $Z^{1/N}$ and dividing the total free energy F by N. This procedure of neglecting the singularity of Z $\to \infty$ is the same as disregarding the singularity of the zeros of $Z^{-1}$. It is concluded that the well-known high-temperature series cannot serve as a standard for judging the putative exact solution of the 3D Ising model. The objections in Perk's Rejoinder [7] (and also in [2,4,5] with respect to the high-temperature series) have been thoroughly disproved.

The author appreciates the support of the National Natural Science Foundation of China (under grant numbers 10674139 and 50831006).

**Additional Replies 1**

After reading Perk's added Comments on arXive:0901.2935v2, I add several additional replies as follows:

It is shown, from his statement about '… the reduced free energy $\beta f = f/k_B T$, not f, near $T = \infty$', that Perk now admits that the replacements $\beta f \to f$ cannot be performed near $T = \infty$. It indicates clearly that $\beta f$ cannot be utilized to discuss singularities at/near $T = \infty$. As discussed already in Part II, one has to face directly the total free energy F to study the singularities at $T = \infty$, because the free energy per site $f = -k_B T \ln \lambda = -k_B \ln \lambda^T$ is then equal to negative infinite, which is inconsistent with the assumption for its definition. Even if one insisted to use f, one must still face all of its singularities at/near $T = \infty$. Although $\lim_{N \to \infty} 1/Z = 0$ occurs also for all finite temperatures (and even at the critical point), the intrinsic characters of singularities of the zero at infinite temperature are quite different from those at finite temperatures. It is seen from the formula of partition function $Z = \exp(-N\beta f)$ that the singularities at finite temperatures originate only from $N \to \infty$, whereas those at/near infinite temperature are much stronger because of the presence of two kinds of singularities ($N \to \infty$ and $T = \infty$). Therefore, singularities at/near infinite temperature cannot be disregarded by the normal process of removing the singularity at finite temperatures.

Yang and Lee did not pay their special attention on singularities at/near $T = \infty$ in their papers, but it does not mean that such singularities are of no physical significance. These singularities are very important for the 3D Ising model. As Perk stated, series and analytical determinations must be starting from near $T = \infty$, not at T

= ∞. This indicates a fact that there is a gap between T → ∞ and T = ∞. How does high-temperature series pass through this gap (with strong singularities, actually, like a 'black hole') from the state at T = ∞? How can analytical determinations from finite temperatures to approach (but never touch) infinite temperature guarantee analyticity of the high-temperature series obtained by accounting deviations from the T = ∞ state?

Next, I point out a fact that Yang and Lee in their papers discussed zeros of Z for evaluating singularities of the thermodynamic consequences: $\lim_{V \to \infty} \frac{1}{V} \ln Z$, $\lim_{V \to \infty} \frac{1}{V} \frac{\partial}{\partial \ln z}(\ln Z)$, $\lim_{V \to \infty} \frac{1}{V} \left(\frac{\partial}{\partial \ln z}\right)^2 (\ln Z)$, … It is clearly seen that according to the Yang – Lee Theorem, one has to face singularities of the logarithmic function lnZ. As illustrated in Part II, one should face also $\ln Z^{-1}$, because considerable interest should be paid to both of them. Actually, if one would always try to conceal singularities of $\ln Z^{-1}$ by mathematical tricks, one would find similar tricks to remove singularities of lnZ also; or, from another angle of view, one would in principle provide with some zeros of lnZ, which could be concealed by the similar procedures, to violate the Yang – Lee Theorem. So, clearly, it is self-contradictory in Perk's Comments.

Finally, no matter analyticity of the reduced free energy has been proved by how many papers contributed by how many groups in how many countries, it does not change a fact that such proofs have never touched T = ∞, where is the point at issue. Agreements between series expansion and other numerical work can tell nothing on the issue debated in this exchange.

**Additional Replies 2**

I add additional replies to Perk's added Comments 2 & 3 (arXiv:0901.2935v3 & v4) on arXive:0901.2935v2:

Although "solving the three-dimensional (3D) Ising model" is a well-known problem, it is set up within the framework of standard equilibrium statistical mechanics with the *ergodic hypothesis* lacking its proof under fairly general conditions. So, using the ensemble average to replace the time average is still questionable, specially, in the 3D Ising model. It is possible that one has to deal with the 3D Ising model within a (3+1)-dimensional framework. However, "black holes" in my last replies is just a metaphor.

The main context of Perk's added Comments 2 (arXiv:0901.2935v3) still focuses on using $\beta f$ for discussing singularities at $T = \infty$, which has already been proved to be invalid. I just add an additional comment: using the basic formulae in thermodynamics, $F = U - TS$, one would get $\beta f = -S/N$ at $T = \infty$, which makes no sense since the entropy is defined for the whole system, not for per site. Furthermore, one would face a problem that at which temperature one should change his interest from $\beta f$ to $F$. The only thing new in Perk's added Comments 2 is about the condition for going to higher dimension. It is my understanding that, at the starting point, the integrand should be performed in four dimensions, since one needs to take the time average by the integrand in the fourth dimension.

As mentioned before, Lebowitz and Penrose indicated clearly in p. 102 of their paper that there is no general reason to expect a series expansion of *p* or *n* in powers of β to converge, since β = 0 lies at the boundary of the region E of (β, z) space. Their proof includes β = 0 only for hard-core potential in section II of their paper, not for the Ising model discussed in other sections. Lebowitz and Penrose at the end of the section II used a word of 'implies' as referred to Gallavotti et al.'s work. However, actually, although Gallavotti et al. proved that the radius of convergence is greater than zero, but once again their proof does not touch β = 0, since the inequality just above (1), i.e., $\sum_{\substack{T \cap X - \phi \\ T \neq \phi}} |K_{\beta\phi'}(X,T)| \leq [\exp(e^{\beta\|\phi'\|} - 1) - 1]$, is invalid for β = 0. In my opinion, one does not need to take lengthy proof in Perk's added Comments 3 (arXiv:0901.2935v4) to prove analyticity of βf and correlation function at β = 0, since βf = - ln 2 at this point for the Ising model, which is of course analyticity. But the key issue here is βf ≠ f ≠ F at β = 0.

I emphasize that the well-known high-temperature and low-temperature series do not take into account the global effects of 'internal factors', but only the local effects, in the transfer matrix, which cannot be 'exact' approaches at finite temperatures. Actually, the former can be exact only a point at/near β = 0. Furthermore, it has not been rigorously established that there is a unique critical point for the 3D Ising model, which provides an opportunity for a transition at/near β = 0. The change of all the interaction energies from zero to non-zero at/near β = 0 results in an intrinsic change of the geometrical (topologic) structure in the 3D Ising system.

In conclusion, the conjectures have not been disproved, and they still open to rigorous prove.

I am grateful to Prof. Dr. Jacques H.H. Perk for discussions via e-mails.